\documentclass[12pt,a4paper]{article}

\usepackage{epsfig}
\usepackage{amsmath,amsfonts,amssymb}
\usepackage{cite}
\usepackage{colortbl}
\usepackage{pifont}

\providecommand{\openone}{\leavevmode\hbox{\small1\kern-3.8pt\normalsize1}}

\newcommand{\gm}{\gamma^\mu}

\newcommand{\smn}{\sigma^{\mu \nu}}

\newcommand{\xctl}{X_{qt}^L}
\newcommand{\xctr}{X_{qt}^R}
\newcommand{\kl}{\kappa_{qt}^L}
\newcommand{\kr}{\kappa_{qt}^R}
\newcommand{\lL}{\lambda_{qt}^L}
\newcommand{\lR}{\lambda_{qt}^R}

\begin{document}

\begin{center}
\begin{Large}
{\bf Ultraboosted $Zt$ and $\gamma t$ production \\[2mm] at the HL-LHC and FCC-hh}
\end{Large}

\vspace{0.5cm}
J.~A.~Aguilar--Saavedra \\
\begin{small}
{ Departamento de F\'{\i}sica Te\'orica y del Cosmos, 
Universidad de Granada, \\ E-18071 Granada, Spain} \\ 

\end{small}
\end{center}

\begin{abstract}
Searches for anomalous $Zt$ and $\gamma t$ production provide an excellent probe of flavour-changing top interactions when the energies considered are very large. In this note we estimate the sensitivity to these interactions at the high luminosity phase of the LHC and a future 100 TeV $pp$ collider (FCC-hh). For the LHC, the expected limits on $t \to uZ / u\gamma $ branching ratios from $Zt$ and $\gamma t$ production will reach the $10^{-5}$ level, one order of magnitude better than the existing projections for $t \to uZ$ from $t \bar t$ production. For the FCC-hh, the limits on $t \to uZ /  u\gamma$ could reach an impressive sensitivity at the $10^{-6}$ level, with limits on $t \to cZ / c\gamma$ at the $10^{-5}$ level.
\end{abstract}

\section{Introduction}

Searches for top flavour-changing neutral (FCN) interactions~\cite{AguilarSaavedra:2004wm} were an important component of the top physics program at the Tevatron and now at the Large Hadron Collider (LHC), and will be pursued at future facilities. Top FCN interactions are extremely weak in the standard model (SM), with branching ratios $\text{Br}(t \to cZ) \sim 10^{-14}$, $\text{Br}(t \to c\gamma) \sim 10^{-14}$, $\text{Br}(t \to cg) \sim 10^{-12}$, $\text{Br}(t \to cH) \sim 10^{-15}$~\cite{Eilam:1990zc,Mele:1998ag,AguilarSaavedra:2002ns}, and values an order of magnitude smaller for the up quark. As such, they offer an excellent window for detectable effects from new physics. Top FCN interactions beyond the SM can be directly probed mainly in two classes of processes: (i) standard $t \bar t$ production with FCN decay of one of the top quarks~\cite{Han:1995pk,Han:1996ep,Han:1996ce,AguilarSaavedra:2000aj}, or (ii) in FCN production of a top quark, followed by a standard decay~\cite{Hosch:1997gz,Han:1998tp,delAguila:1999ac,AguilarSaavedra:2000aj}. At the LHC, the processes in the first class are expected to be more sensitive for $tc$ interactions, while for $tu$ interactions the sensitivity is expected to be comparable in both cases. These projections are confirmed by actual analyses with data. Searching for $t \bar t$ production with FCN decay of one top quark, the CMS Collaboration has obtained the limits $\text{Br}(t \to cZ,uZ) \leq 5 \times 10^{-4}$ with a 95\% confidence level (CL), using the full Run 1 dataset with 19.7 fb$^{-1}$~\cite{Chatrchyan:2013nwa}. A recent combination with anomalous $tZ$ production has improved these limits to $\text{Br}(t \to cZ) \leq 4.9 \times 10^{-4}$, $\text{Br}(t \to uZ) \leq 2.2 \times 10^{-4}$~\cite{Sirunyan:2017kkr}. For top decay processes, the ATLAS Collaboration sets weaker limits, $\text{Br}(t \to cZ,uZ) \leq 8 \times 10^{-4}$~\cite{Aad:2015uza}, while searches for $tZ$ production have not been performed.

At future facilities, limits on top FCN interactions resulting from $t \bar t$ production will not significantly improve over the current ones. For example, the projections from the ATLAS Collaboration for the high-luminosity Large Hadron Collider (HL-LHC) with 3000 fb$^{-1}$ at 14 TeV are $\text{Br}(t \to cZ) \leq 2.3 \times 10^{-4}$, $\text{Br}(t \to uZ) \leq 1.3 \times 10^{-4}$~\cite{ATL-PHYS-PUB-2016-019}. The reason is that, unfortunately, those searches will soon be dominated by systematic uncertainties, and will not benefit from the huge luminosity upgrade. 
On the other hand, single top FCN production processes, when mediated by effective non-renormalisable interactions, lead to distinct kinematical signatures, which allow to perform measurements at high transverse momenta where the SM background is small. These features were already exploited when estimating the LHC sensitivity to top FCN interactions in $Zt$ and $\gamma t$ production~\cite{delAguila:1999kfp}. The Lagrangian for top flavour-changing interactions with the $Z$ boson and photon can be written as~\cite{AguilarSaavedra:2008zc}
\begin{eqnarray}
\mathcal{L}_{Zt} & = & - \frac{g}{2 c_W} \bar q \, \gm \left( \xctl P_L
  + \xctr P_R \right) t\; Z_\mu  \nonumber \\
& & - \frac{g}{2 c_W} \bar q \, \frac{i \smn q_\nu}{M_Z}
  \left( \kl P_L + \kr P_R \right) t\; Z_\mu  + \text{H.c.}  \,, \notag \\
\mathcal{L}_{\gamma t} & = & - e \bar q \, \frac{i \smn q_\nu}{m_t} \left( \lL P_L
+ \lR P_R \right) t\; A_\mu + + \text{H.c.}  \,,
\end{eqnarray}
with $q=u,c$.
Among these, the dipole $\smn$ terms have the high-energy enhancement mentioned before, and limits on them are naively expected to be very stringent in the ultraboosted regime. But, besides this improvement, in the ultraboosted regime there is a drawback: the decay products of the top quark merge into a single jet. In that situation, the main SM background is $Zj$ and $\gamma j$ production, with $j$ a quark or gluon jet, whose cross section is huge, much larger than $ZWj$ and $\gamma Wj$, which would the corresponding irreducible backgrounds if the top decay products could be resolved. Then, it is not obvious that considering the ultraboosted regime for these searches will actually be an advantage, and a detailed assessment of the sensitivity is necessary in order to seriously consider these channels from the experimental point of view. This is the purpose of this note, focusing on the HL-LHC and a future 100 TeV $pp$ circular collider (FCC-hh), and restricting ourselves to $Zt$ and $\gamma t$ production mediated by non-renormalisable couplings.
We consider in turn the sensitivity of the HL-LHC in section~\ref{sec:2} and of the FCC-hh in section~\ref{sec:3}. In section 4 we summarise our results and discuss possible improvements to the simple analyses outlined here.


\section{Sensitivity at the HL-LHC}
\label{sec:2}

Our analysis is performed with a fast simulation of the (upgraded) CMS detector. The generation of the $Zt$ and $\gamma t$ signals is done with {\scshape Protos}~\cite{AguilarSaavedra:2010rx,protos}. We set the anomalous couplings $\kl = 0.01$, $\kr = 0$, $\lL = 0.01$, $\lR = 0$. (The specific choice of left-handed chirality for the interactions hardly affects the results.) The showering and hadronisation is performed by {\scshape Pythia8}~\cite{Sjostrand:2007gs} and the detector simulation by {\scshape Delphes 3.4}~\cite{deFavereau:2013fsa}. The jets are reconstructed with {\scshape FastJet}~\cite{Cacciari:2011ma} using the anti-$k_T$ algorithm~\cite{Cacciari:2008gp} and trimmed~\cite{Krohn:2009th} using the parameters $R=0.2$, $f_\text{cut} = 0.05$. At LHC energies we require for event selection the presence of a large-radius $R=1.0$ jet $J$, with transverse momentum $p_T^J > 600$ GeV and pseudorapidity $|\eta| < 2.5$. In $Zt$ production we require the presence of two leptons $\ell=e,\mu$ of the same flavour, with transverse momentum $p_T^\ell > 200$ GeV and pseudo-rapidity $|\eta| < 2.4$, having an invariant mass  $m_{\ell \ell}$ in the interval $[60,120]$ GeV. In $\gamma t$ production we require a photon with $p_T^\gamma > 400$ GeV. In addition, we require that the lego-plot separation $\Delta R$ between the jet and the reconstructed $Z$ boson or photon is larger than 1. We divide the event sample into `semileptonic' and `hadronic' channels depending on whether a hard lepton is found within the jet:
\begin{itemize}
\item[(i)] semileptonic: a lepton (the leading lepton in the event in $\gamma t$, the third one in $Zt$) is found within $\Delta R = 0.3$ distance of the jet, and with transverse momentum fraction $z \equiv p_T^\ell / p_T^J > 0.1$;
\item[(ii)] hadronic: if the above condition is not fullfilled.
\end{itemize}
The choice of the cuts on $z$ fraction and $\Delta R$ distance are good to obtain a large rejection of the leading $Zj$ and $\gamma j$ backgrounds, in which leptons are produced from cascade decays within the jets, but is not optimised. 

While for the hadronic channel the main backgrounds are, by far,  $Zj$ and $\gamma j$ production, for the semileptonic channel $Zb / \gamma b$ and $ZWj / \gamma Wj$ are important too. These backgrounds are generated with {\scshape MadGraph5}~\cite{Alwall:2014hca}. All the signals and backgrounds are generated by slicing the phase space in intervals of top or jet transverse momenta, so that the region of large $p_T$, with cross sections two orders of magnitude smaller than the lowest $p_T$ range considered, is populated with sufficient statistics. In all cases, top quark and anti-quark production is summed.

\begin{figure}[htb]
\begin{center}
\begin{tabular}{cc}
\includegraphics[height=5cm,clip=]{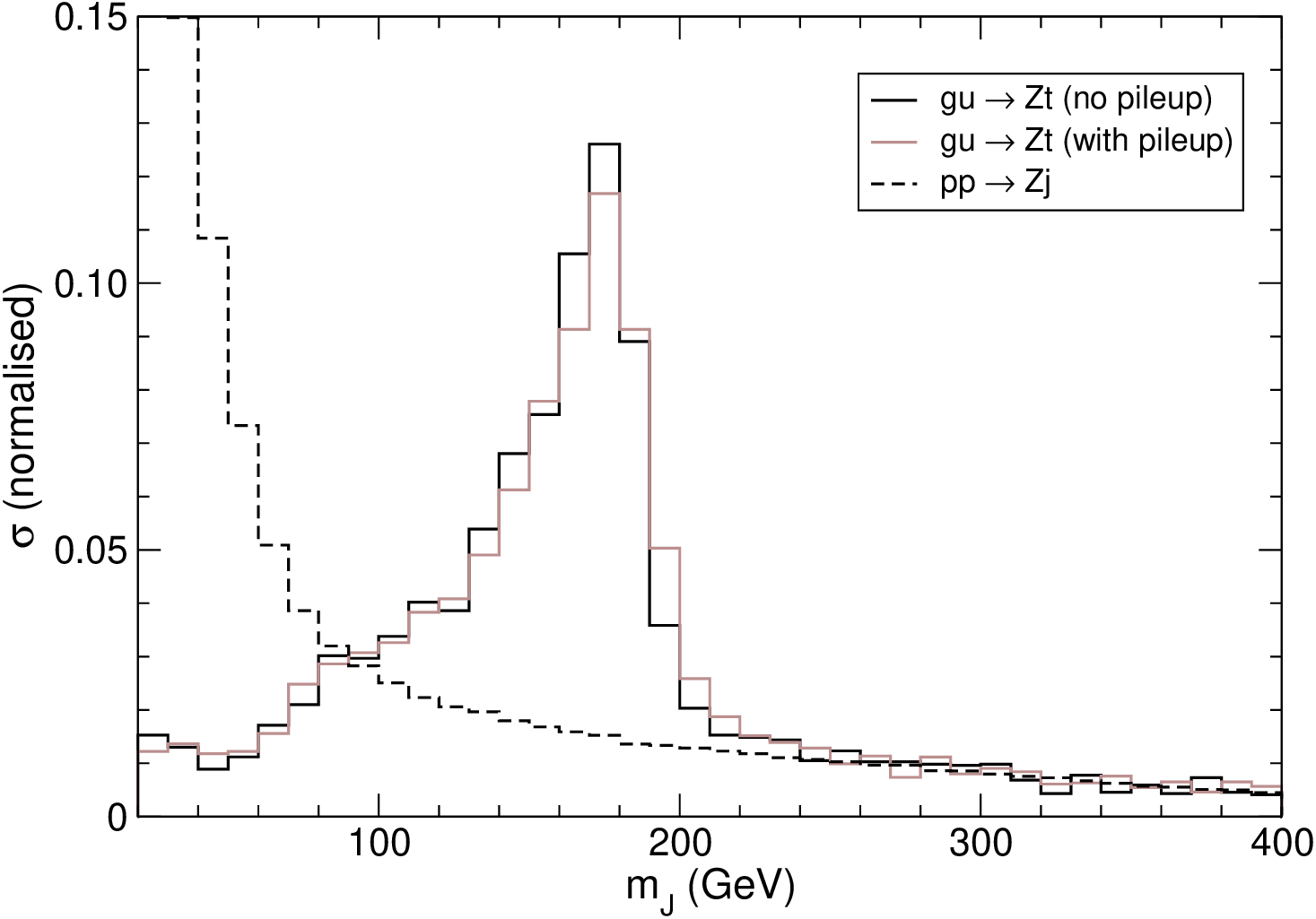} &
\includegraphics[height=5cm,clip=]{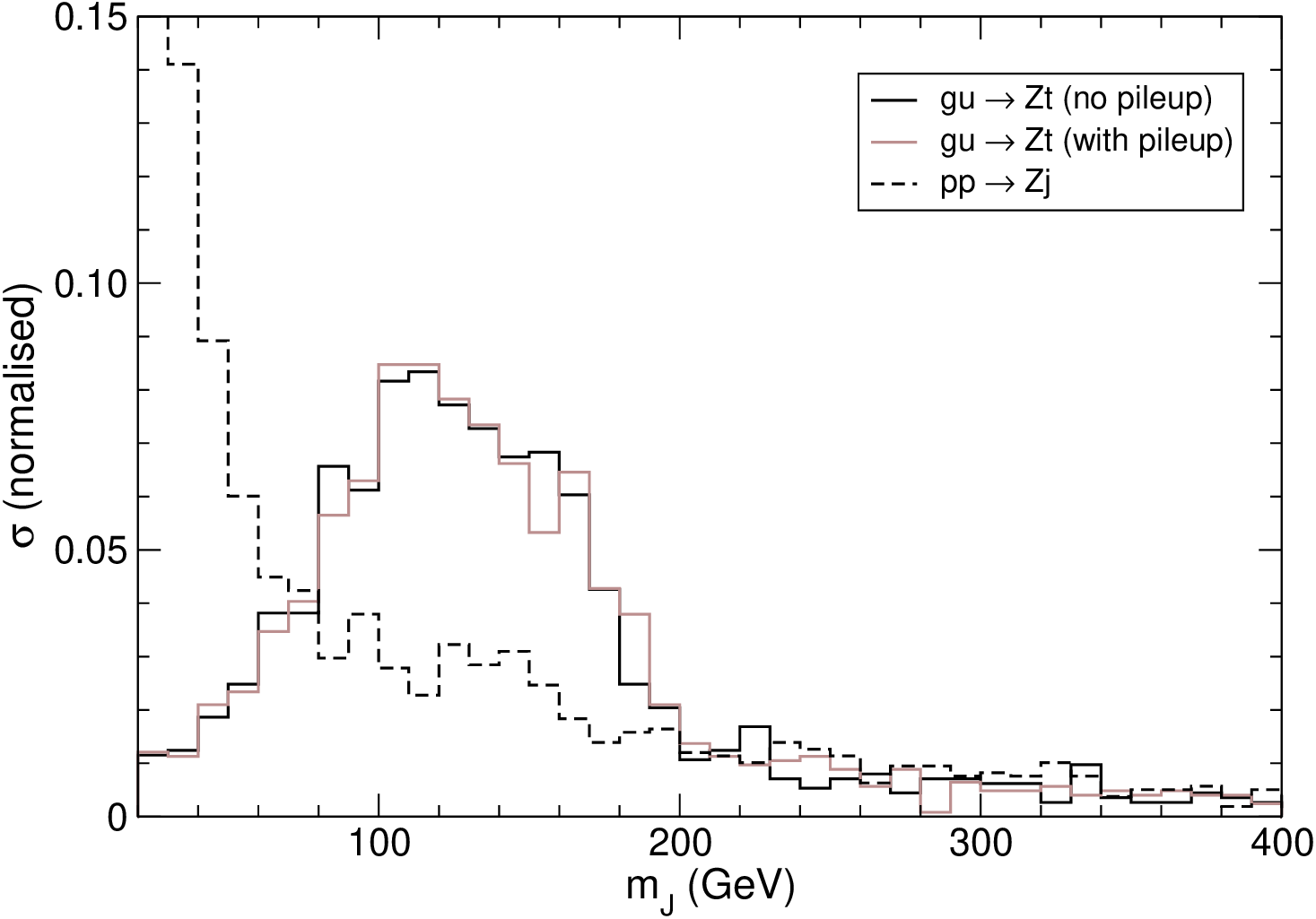}
 \end{tabular}
\caption{Jet mass distribution for the $Zt$ signal (with and without pileup) and the $Zj$ background. Left: hadronic channel. Right: semileptonic channel.}
 \label{fig:mJ}
\end{center}
\end{figure}

Pileup events are a major issue at the HL-LHC, and unfortunately the simulation of the signals and backgrounds with enough statistics and the expected 140 pileup events per beam crossing is computationally very CPU and storage demanding (around 20 TB of disk space for the Monte Carlo statistics of several million events used in our simulations). To overcome this technical difficulty, we base our signal and background discrimination on pileup-robust observables, such as the jet $p_T$ and mass, and do not simulate pileup events to obtain our estimates. For illustration, we show in Fig.~\ref{fig:mJ} the jet mass distribution after trimming for $g u \to Zt$ events, with and without pileup, for a sample of $2 \times 10^4$ events generated with top transverse momentum between 1.2 and 1.4 TeV at the parton level. For comparison we also show the jet mass distribution for a sample of $8 \times 10^5$ $Zj$ events with jet $p_T$ between 1.2 and 1.4 TeV, still at the parton level. The left panel corresponds to the hadronic channel, in which the mass is well reconstructed, and the right panel to the semileptonic channel, in which the missing neutrino does not contribute to the jet mass. 
One can observe that the trimmed jet mass is not significantly affected by pileup. For our event selection, in the hadronic channel we further require a jet mass $m_J \in [150,200]$ GeV, and in the semileptonic channel $m_J \in [100,175]$ GeV.

\begin{figure}[htb]
\begin{center}
\begin{tabular}{cc}
\includegraphics[height=5cm,clip=]{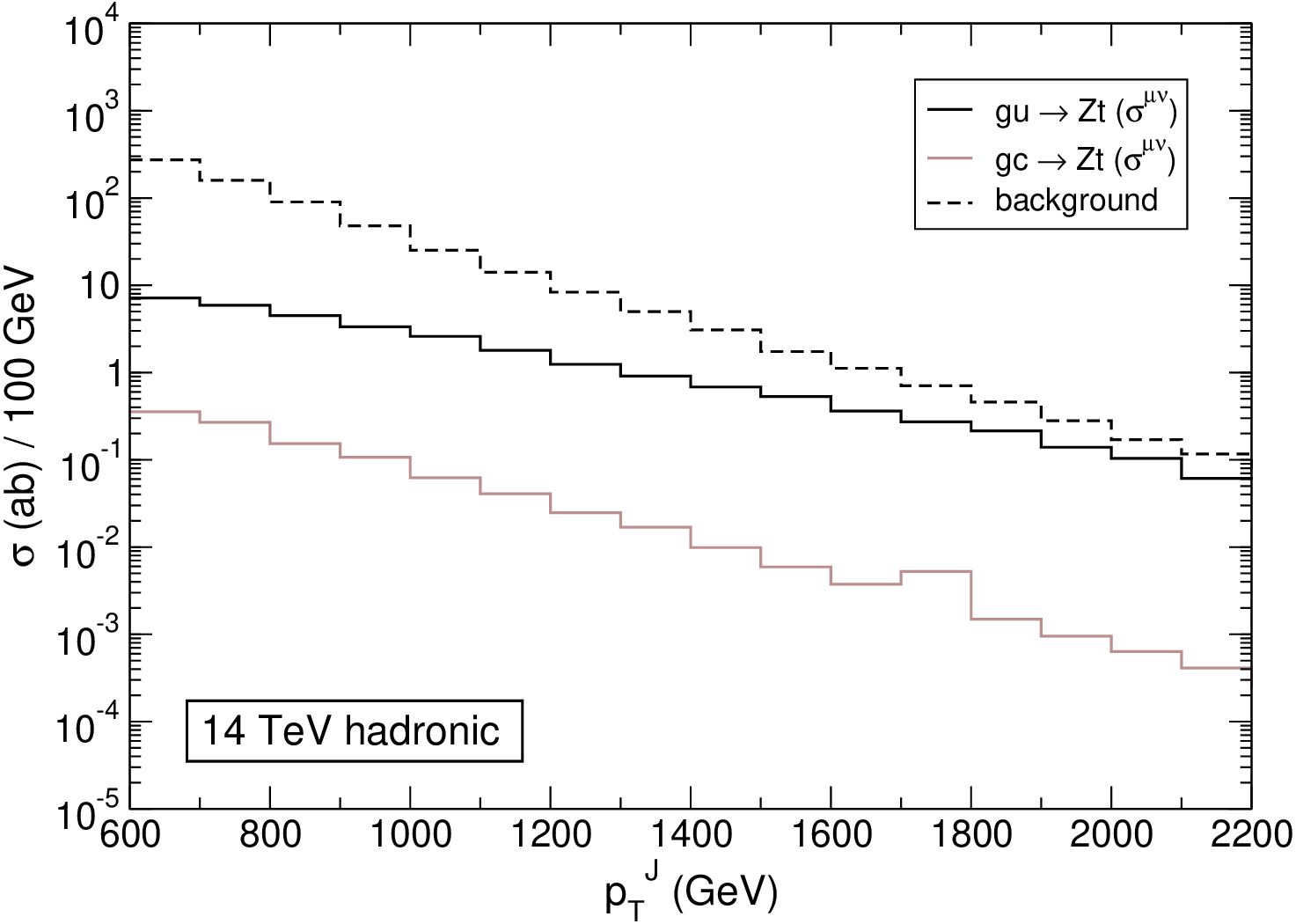} &
\includegraphics[height=5cm,clip=]{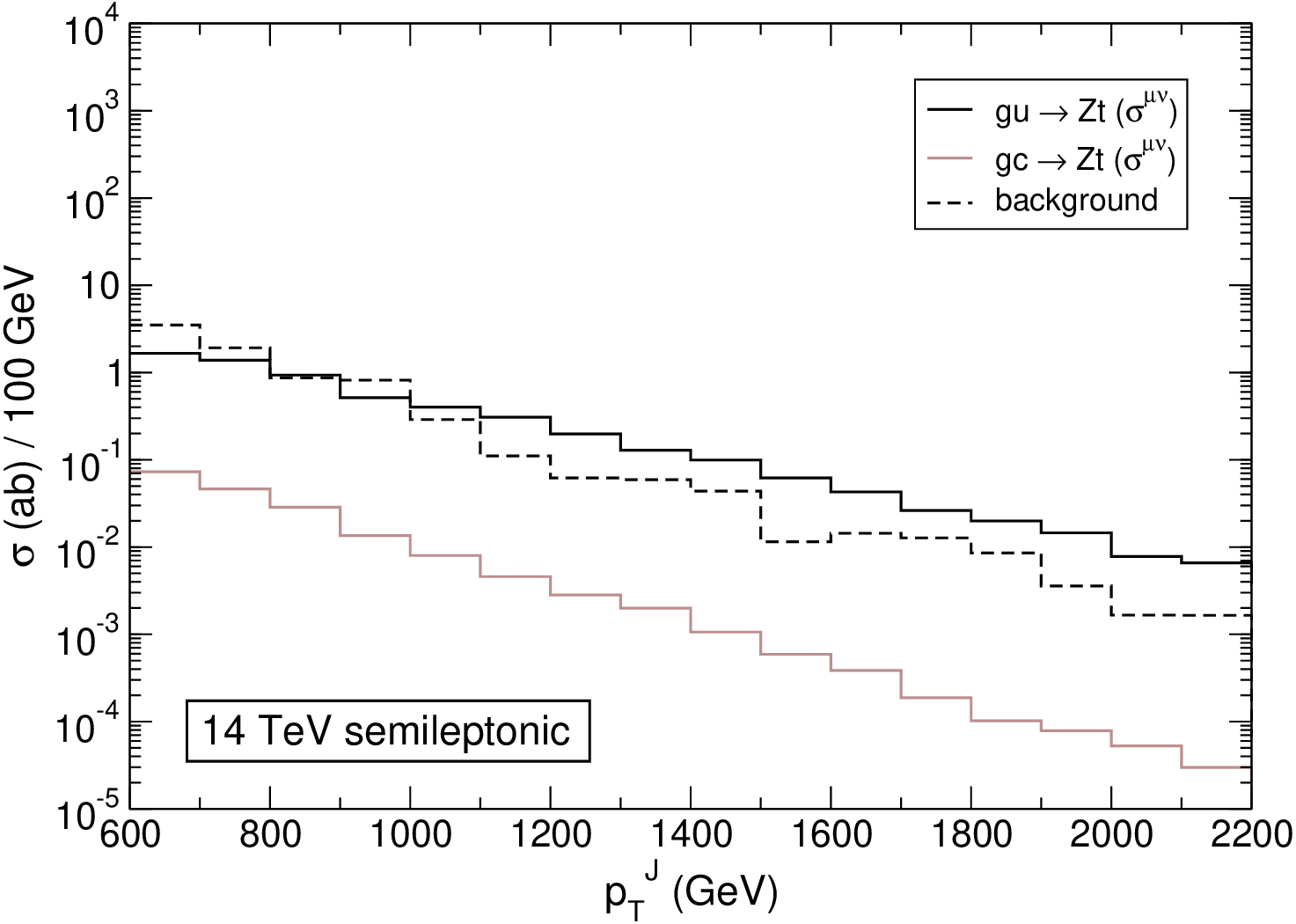} \\
\includegraphics[height=5cm,clip=]{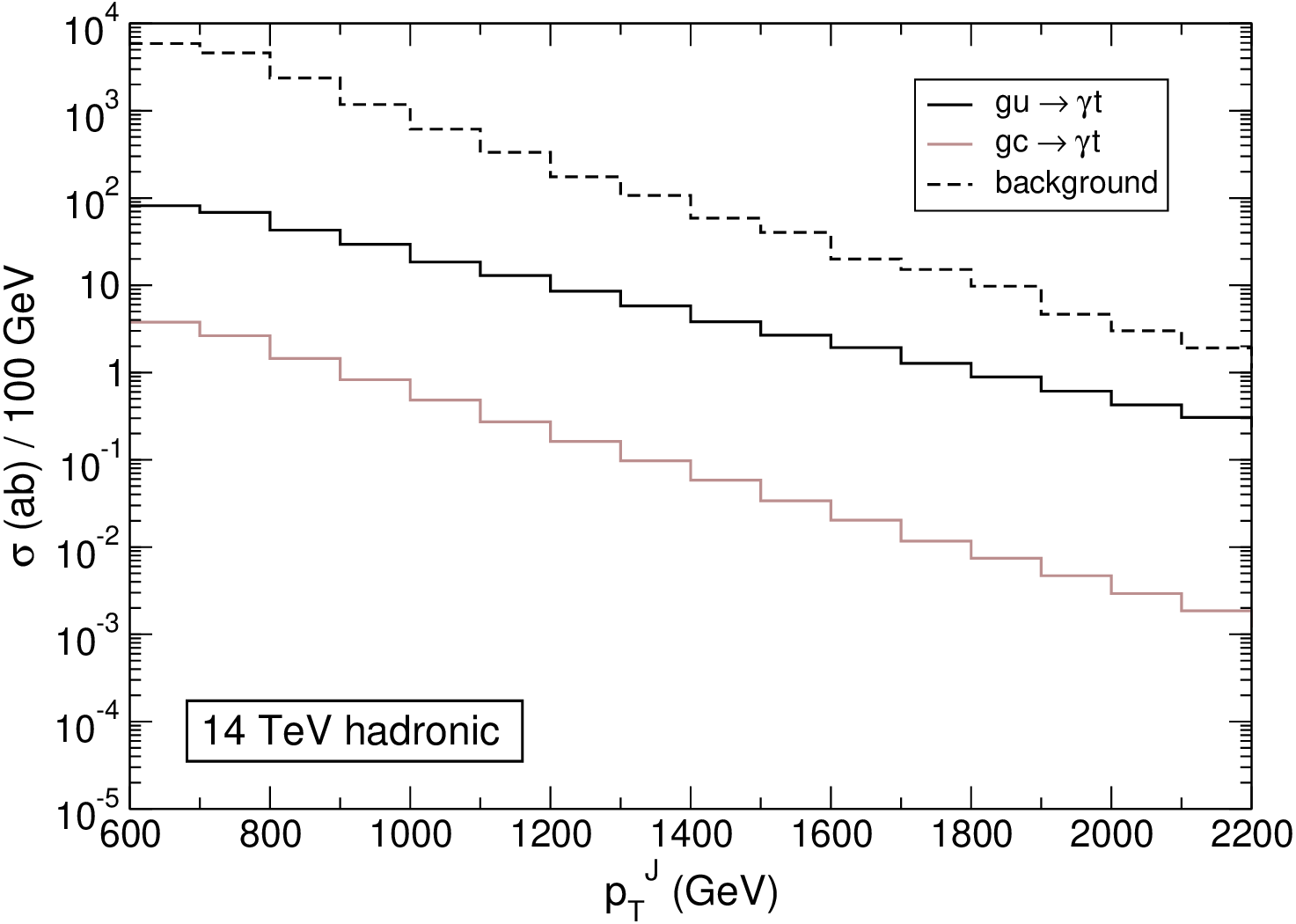} &
\includegraphics[height=5cm,clip=]{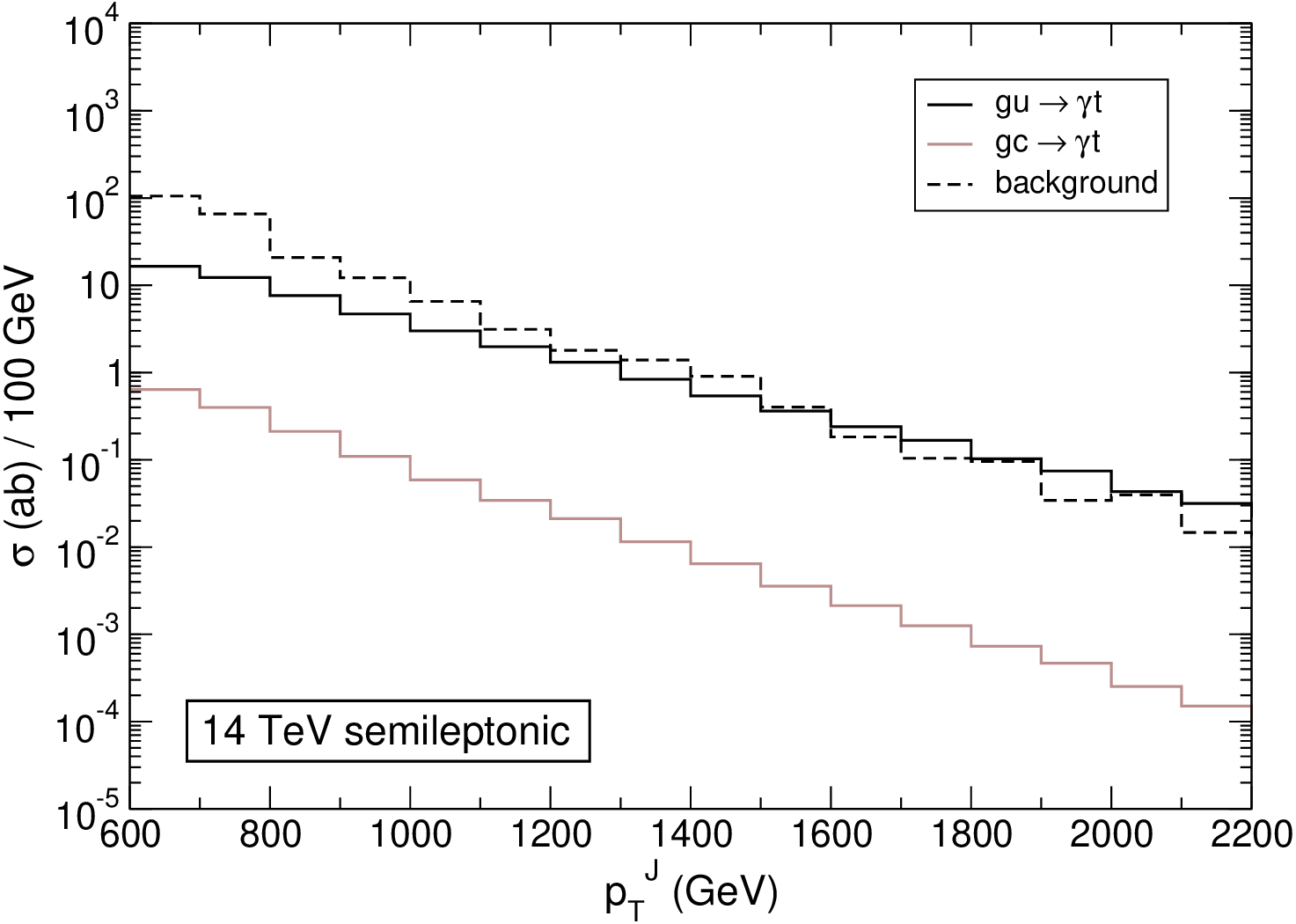}
 \end{tabular}
\caption{Leading jet transverse momentum distribution for the $Zt$ and $\gamma t$ signals and their background at the HL-LHC, in the hadronic and semileptonic channels.}
 \label{fig:pT14}
\end{center}
\end{figure}

The transverse momentum distribution of the hardest jet in the event, deemed to correspond to the top quark in the case of the signals, is presented in Fig.~\ref{fig:pT14}, for the $Zt$ (top) and $\gamma t$ (bottom) signals and their backgrounds, in the hadronic (left) and semileptonic (right) channels. One can clearly see that for the $u$-initiated signals the slope of the distributions is milder than for the background~\cite{Mangano:2016jyj}, showing the advantage of considering ultraboosted top quarks. On the other hand, for $c$-initiated processes the slope is similar and the benefit of going to high $p_T^J$ is moderate.

The advantage of requiring a hard lepton within the jet~\cite{Brust:2014gia,Aguilar-Saavedra:2014iga} to enhance the signal significance is also manifest: the leading $Zj/\gamma j$ backgrounds are suppressed by two orders of magnitude. For illustration, we collect in Table~\ref{tab:sig14} the cross sections of the different signal and background processes considered, after the event selection. For comparison, the cross sections for SM $tZj$ production with the event selection considered are 2.6 (0.36) ab in the hadronic (semileptonic) channel. 

\begin{table}[t]
\begin{center}
\begin{tabular}{ccccccc}
& hadronic & semileptonic & & & hadronic & semileptonic \\
$gu \to Zt$ & $30$ & $5.8$ &&
$gu \to \gamma t$ & $280$ & $50$ \\
$gc \to Zt$ & $1.1$ & $0.18$ &&
$gc \to \gamma t$ & $10$ & $1.5$ \\
$Z j$ & $604$ & 4.8 & \quad &
$\gamma j$ & $1.5 \times 10^4$ & 193 \\
$Z b$ & $15$ & 2.3 & \quad &
$\gamma b$ & $81$ & $17$ \\
$Z W j$ & $13$ & $0.66$ & \quad &
$\gamma W j$ & $206$ & $9$ 
\end{tabular}
\end{center}
\caption{Cross sections (in ab) for the different signal and background processes at the HL-LHC, after the event selection. For the signals, we take $\kl = 0.01$, $\lL = 0.01$.}
\label{tab:sig14}
\end{table}

\begin{figure}[t]
\begin{center}
\begin{tabular}{cc}
\includegraphics[height=5cm,clip=]{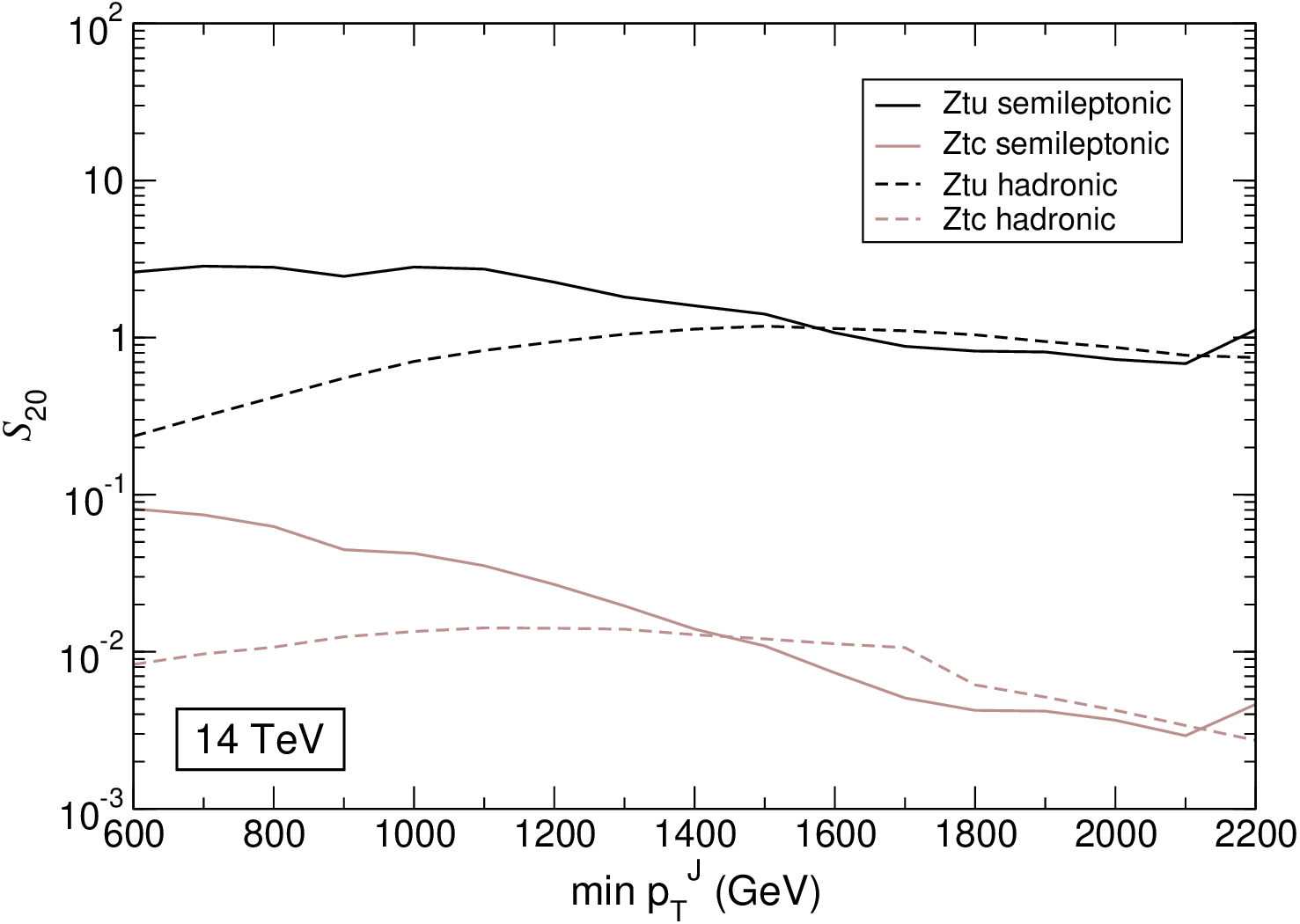} &
\includegraphics[height=5cm,clip=]{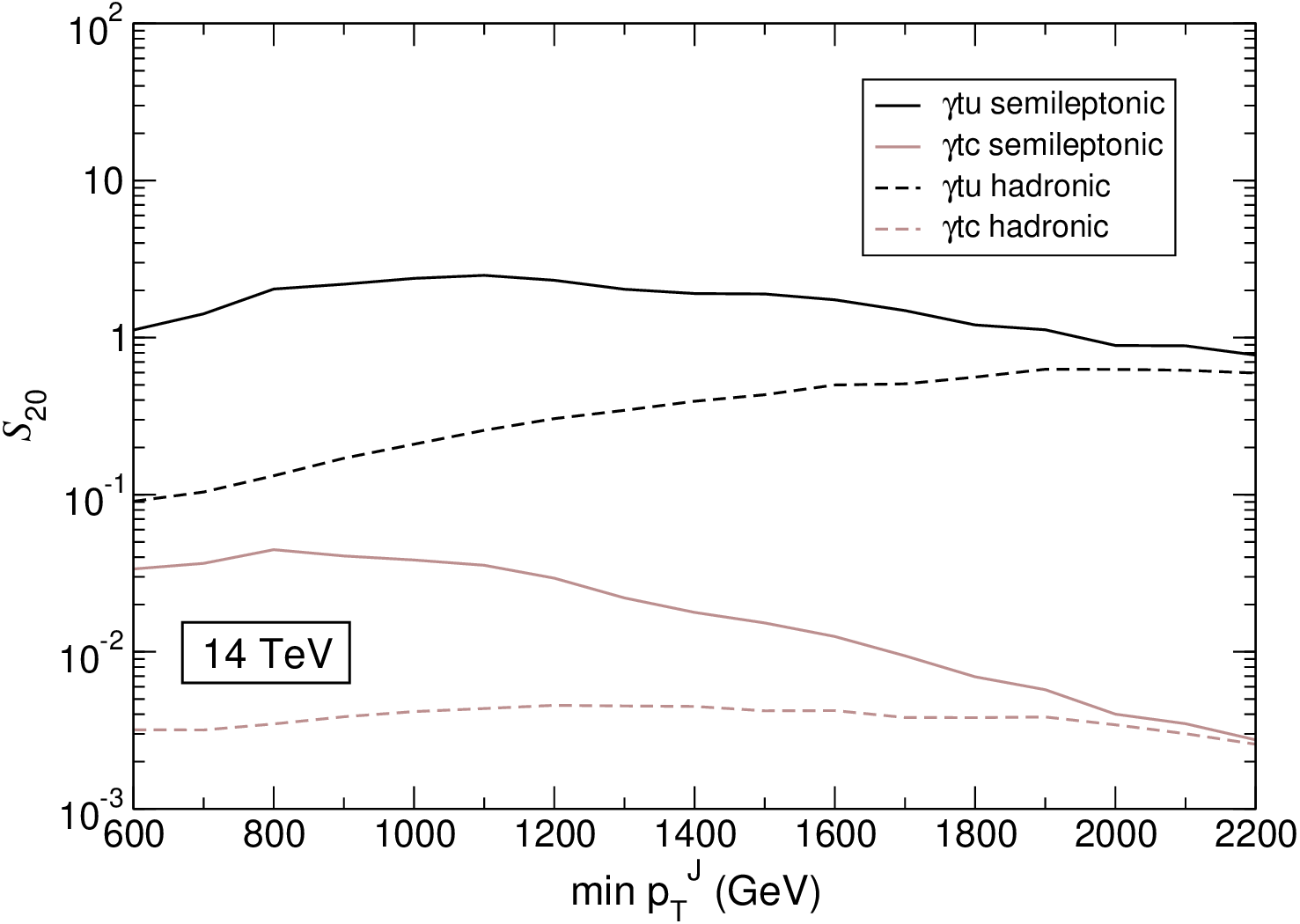}
 \end{tabular}
\caption{Value of the estimator $\mathcal{S}_{20}$ in Eq.~(\ref{ec:s20}) as a function of the cut on minimum $p_T^J$, for the HL-LHC.}
 \label{fig:signif14}
\end{center}
\end{figure}

To estimate the upper limits that could be placed on a possible FCN signal, we use the estimator
\begin{equation}
\mathcal{S}_{20} = \frac{S}{\sqrt{B + (0.2 B)^2}} \,,
\label{ec:s20}
\end{equation} 
with $S$ and $B$ the number of signal and background events, respectively. When the number of background events is sufficiently large, the denominator of this quantity gives the background uncertainty as the statistical one ($\sqrt B$), plus a 20\% systematic uncertainty, summed in quadrature.  We plot $\mathcal{S}_{20}$, assuming a luminosity of 3 ab$^{-1}$, in Fig.~\ref{fig:signif14}, as a function of the lower cut on the leading jet transverse momentum, denoted as min $p_T^J$.
Of course, for very high values of this cut, the number of background events is small and Poisson statistics must be used, instead of the Gaussian approximation in the definition of $\mathcal{S}_{20}$. However, we have checked that for the values of the min $p_T^J$ cut where $\mathcal{S}_{20}$ is near its maximum, the number of background events is sufficiently large, as it can also be seen from the distributions in Fig.~\ref{fig:pT14}. Using $\mathcal{S}_{20}$ as estimator, and translating the upper limits on $\kl$ and $\lL$ into top decay branching ratios, we obtain at the 95\% CL
\begin{align}
& \text{Br}(t \to uZ) < 4.1 \times 10^{-5} \quad (1.1 \times 10^{-4}) \notag \,, \\
& \text{Br}(t \to cZ) < 1.6 \times 10^{-3} \quad (9.6 \times 10^{-3}) \notag \,, \\
& \text{Br}(t \to u\gamma) < 1.8 \times 10^{-5} \quad (5.3 \times 10^{-5}) \,, \notag \\
& \text{Br}(t \to c\gamma) < 6.1 \times 10^{-4} \quad (5.2 \times 10^{-3}) \,,
\end{align}
with the numbers between parentheses corresponding to the hadronic channel.
While the limits for top FCN couplings with the charm quark are relatively weak as expected, those with the up quark are very stringent. For $t \to Zu$, a comparison with projected limits from $t \bar t$ production in Ref.~\cite{ATL-PHYS-PUB-2016-019} is possible. Both the hadronic and semilteptonic channel in $Zt$ production improve over the expectations from top decays, up to an order of magnitude in the latter case.


\section{Sensitivity at the FCC-hh}
\label{sec:3}

\begin{figure}[htb]
\begin{center}
\begin{tabular}{cc}
\includegraphics[height=5cm,clip=]{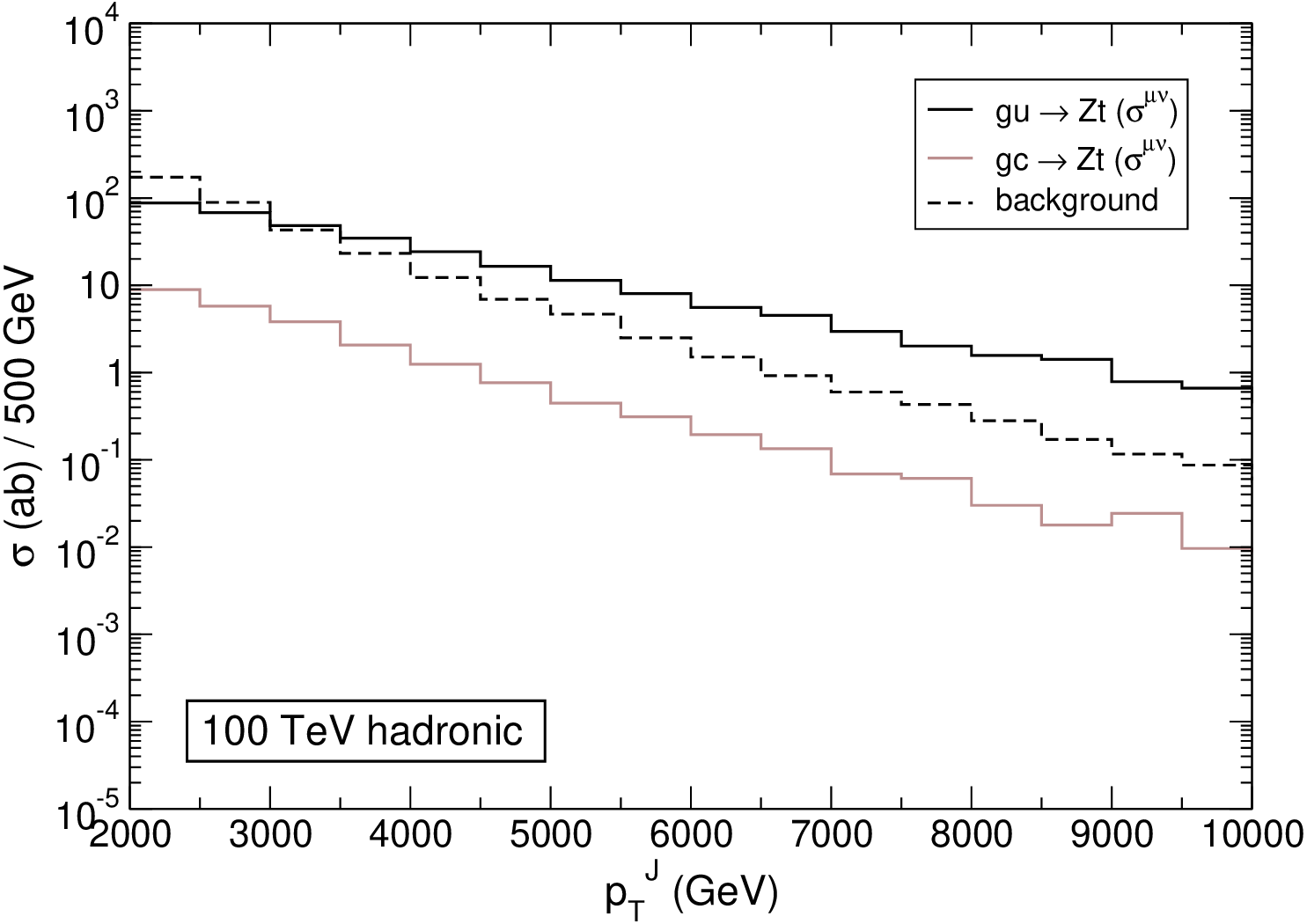} &
\includegraphics[height=5cm,clip=]{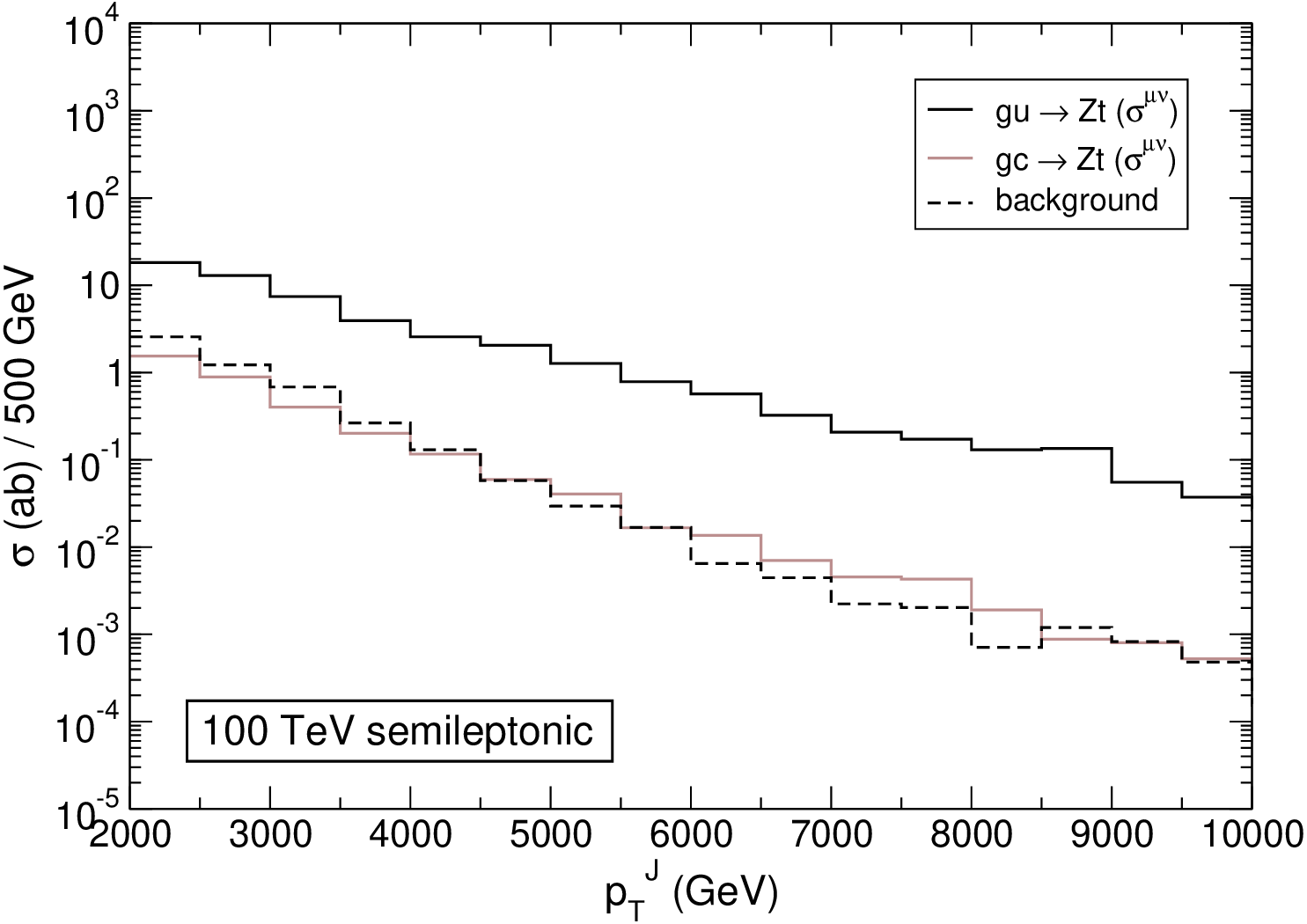} \\
\includegraphics[height=5cm,clip=]{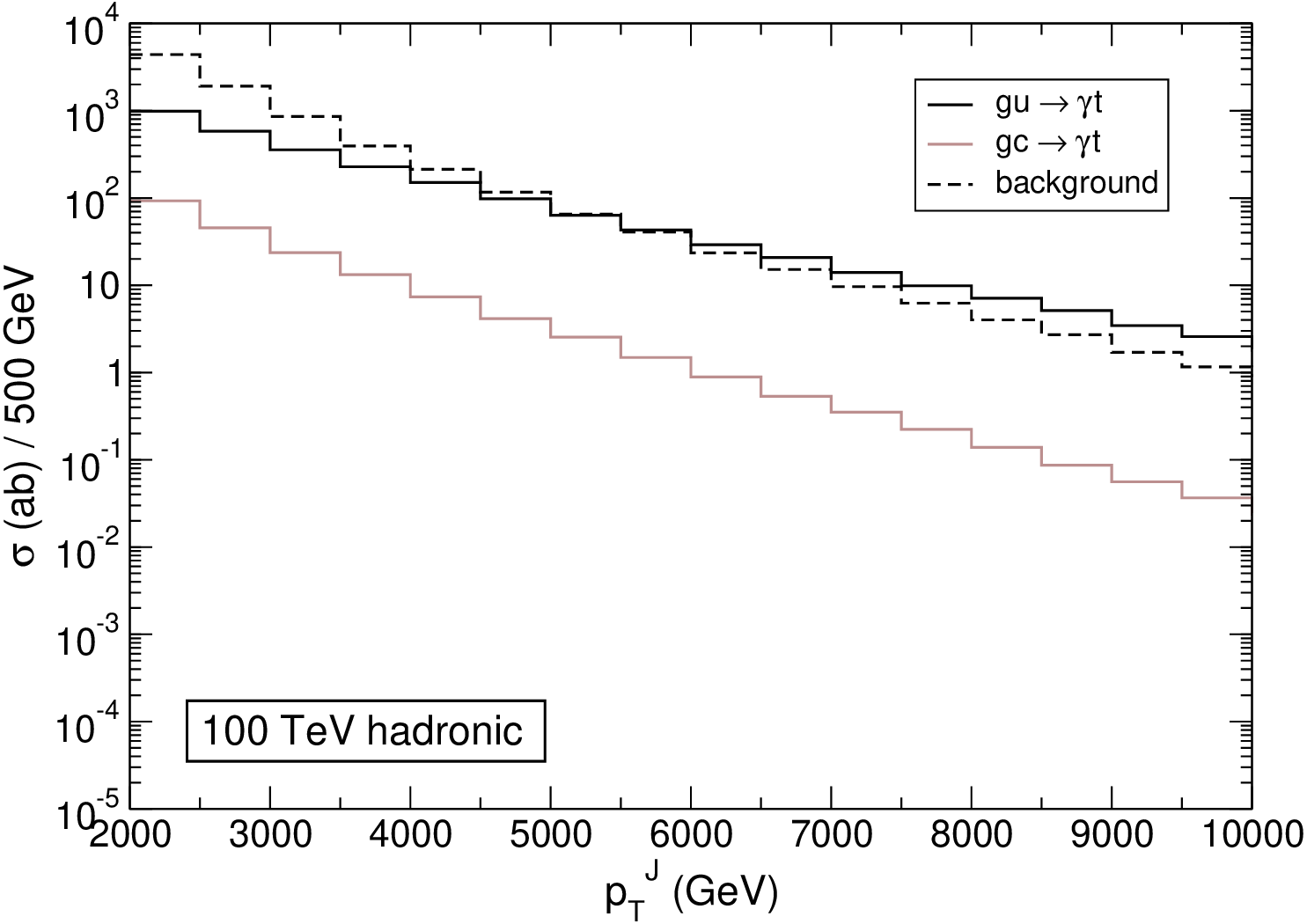} &
\includegraphics[height=5cm,clip=]{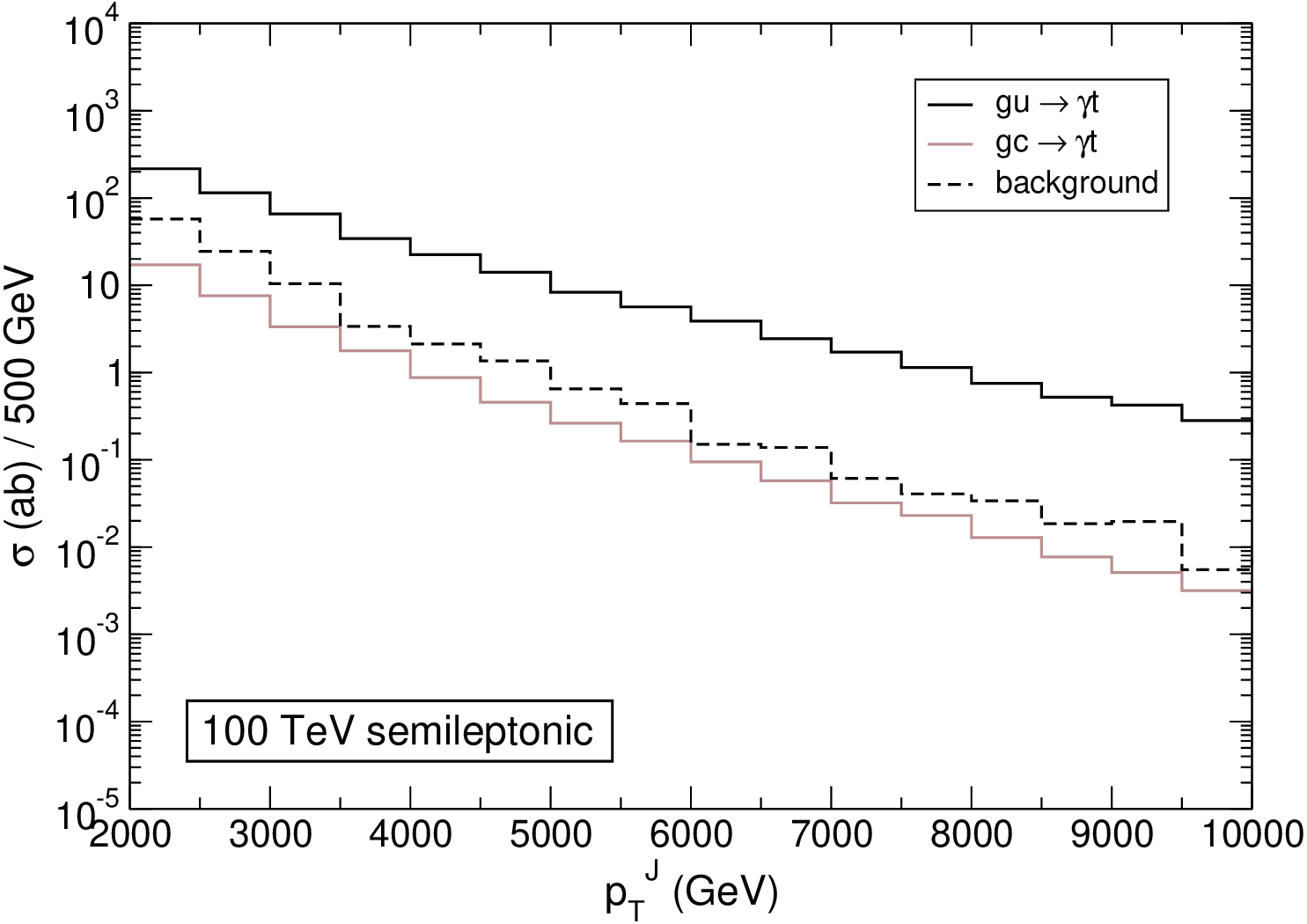}
 \end{tabular}
\caption{Leading jet transverse momentum distribution for the $Zt$ and $\gamma t$ signals and their background at the FCC-hh, in the hadronic and semileptonic channels.}
 \label{fig:pT100}
\end{center}
\end{figure}

We perform this analysis with a fast simulation of a future FCC-hh detector. 
The event selection is analogous to the one described in the previous section, but using jets of radius $R=0.4$ 
and raising the transverse momentum thresholds to $p_T > 2000$ GeV for the leading jet, $p_T > 500$ GeV for charged leptons and $p_T > 1000$ GeV for the photon. The semileptonic channel is defined taking $\Delta R = 0.2$ and keeping $z > 0.1$. The corresponding distributions for the transverse momentum of the leading jet are given in Fig.~\ref{fig:pT100}.
The behaviour is qualitatively the same as seen in the previous section. The cross sections for the different processes after event selection are given in Table~\ref{tab:sig100}.
\begin{table}[htb]
\begin{center}
\begin{tabular}{ccccccc}
& hadronic & semileptonic & & & hadronic & semileptonic \\
$gu \to Zt$ & $320$ & $51$ &&
$gu \to \gamma t$ & $2500$ & $490$ \\
$gc \to Zt$ & $24$ & $3.3$ &&
$gc \to \gamma t$ & $190$ & $32$ \\
$Z j$ & $320$ & $2.6$ & \quad &
$\gamma j$ & $7700$ & 75 \\
$Z b$ & $13$ & 2.0 & \quad &
$\gamma b$ & $77$ & $13$ \\
$Z W j$ & $24$ & $0.34$ & \quad &
$\gamma W j$ & $340$ & $14$ 
\end{tabular}
\end{center}
\caption{Cross sections (in ab) for the different signal and background processes at the FCC-hh, after the event selection. For the signals, we take $\kl = 0.01$, $\lL = 0.01$.}
\label{tab:sig100}
\end{table}

\begin{figure}[htb]
\begin{center}
\begin{tabular}{cc}
\includegraphics[height=5cm,clip=]{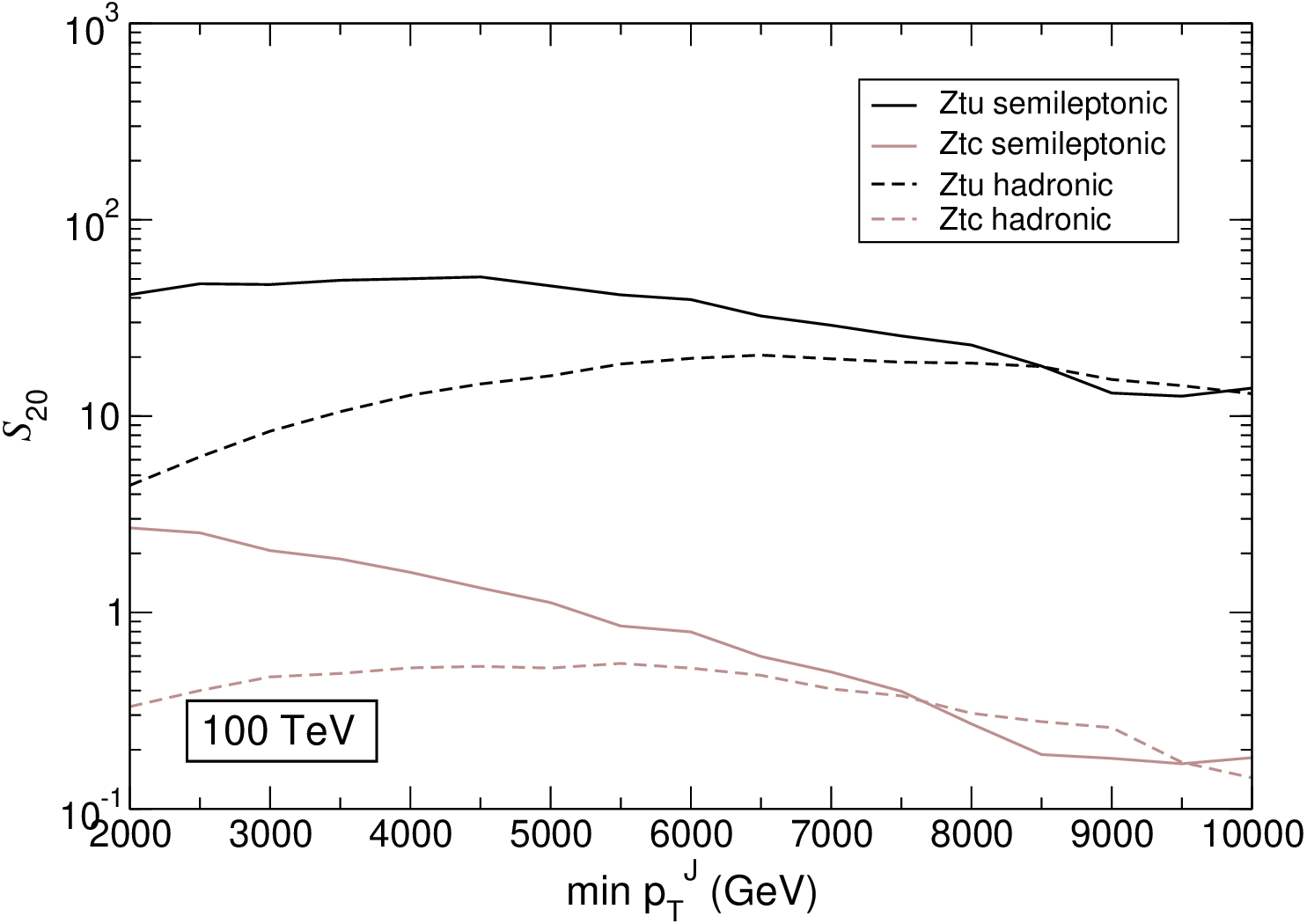} &
\includegraphics[height=5cm,clip=]{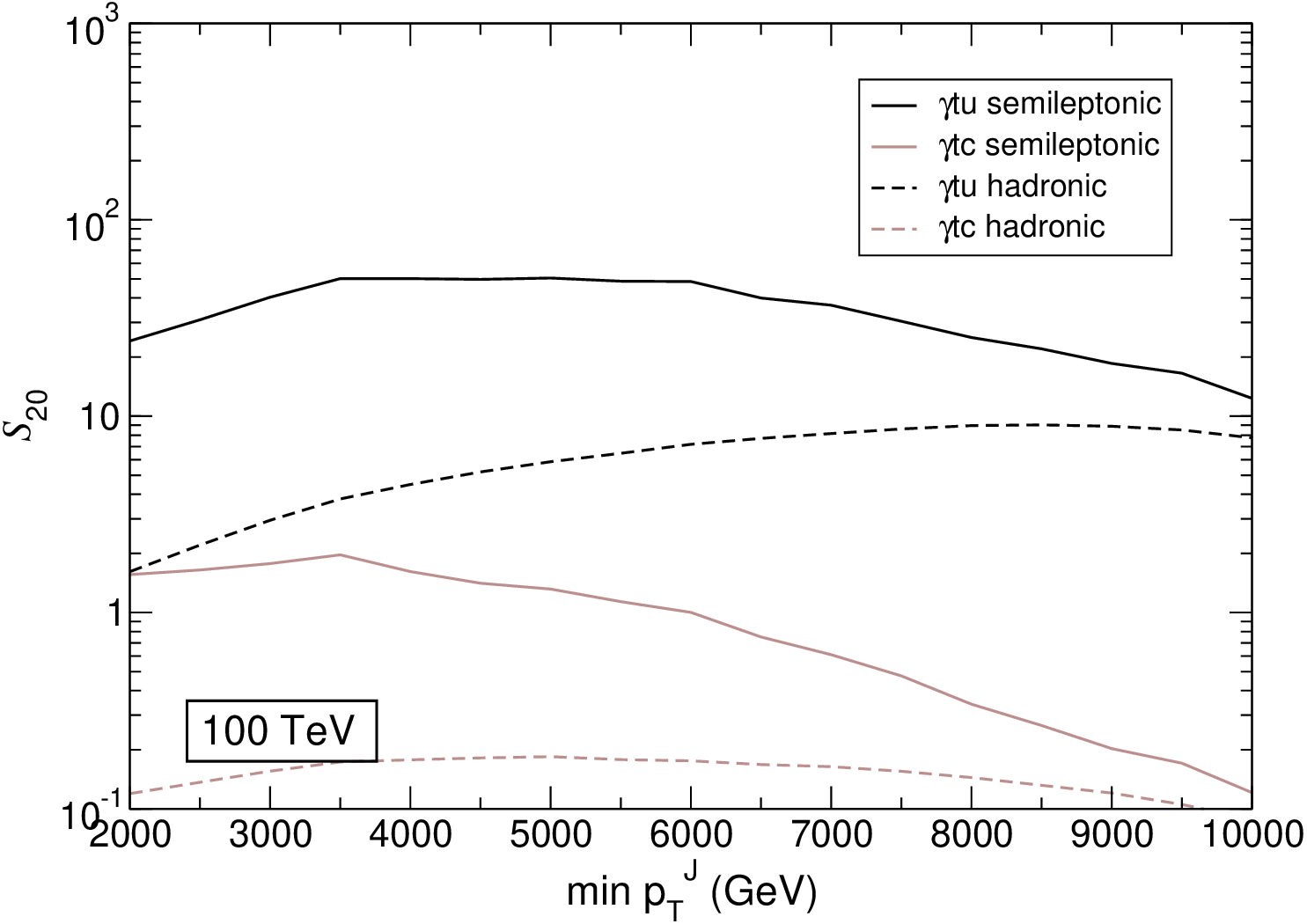}
 \end{tabular}
\caption{Value of the estimator $\mathcal{S}_{20}$ in Eq.~(\ref{ec:s20}) as a function of the cut on minimum $p_T^J$, for the FCC-hh.}
 \label{fig:signif100}
\end{center}
\end{figure}

With the huge energy boost at the FCC-hh, $Zt$ and $\gamma t$ processes allow to obtain competitive limits even on top FCN interactions with the charm quark. We present in Fig.~\ref{fig:signif100} the estimator $\mathcal{S}_{20}$ as a function of the cut on $p_T^J$, assuming a luminosity of 10 ab$^{-1}$. 
The estimated 95\% CL upper limits that could be achieved with that luminosity are quite remarkable, reaching and even surpassing the $10^{-6}$ ballpark in the case of the up quark,
\begin{align}
& \text{Br}(t \to uZ) < 2.7 \times 10^{-6} \quad (6.5 \times 10^{-6}) \notag \,, \\
& \text{Br}(t \to cZ) < 5.0 \times 10^{-5} \quad (2.5 \times 10^{-4}) \notag \,, \\
& \text{Br}(t \to u\gamma) < 9.1 \times 10^{-7} \quad (5.1 \times 10^{-6}) \,, \notag \\
& \text{Br}(t \to c\gamma) < 2.3 \times 10^{-5} \quad (2.5 \times 10^{-4}) \,.
\end{align}

\section{Discussion}
\label{sec:4}

The analysis performed in this note shows that searches for $Zt$ and $\gamma t$ production in the ultraboosted regime will provide competitive limits on top FCN interactions, despite the fact that backgrounds are very large when the top decay products cannot be resolved. We have performed a simple and robust analysis, based on fast detector simulations, in which the variables used to discriminate signal and background are the mass and transverse momentum of the leading jet, which corresponds to the top quark in the two signals under study. And we have used a third variable, the fraction of transverse momentum of an additional lepton within the jet, to distinguish two top quark decay channels, semileptonic and hadronic, with the backgrounds being much smaller for the former.

With the analysis presented here we have found that the expected sensiftivity to $tuZ$ and $tu\gamma$ interactions is excellent. For $tuZ$ interactions for which a comparison with other analyses is possible, we find that the projected limits on the $t \to uZ$ branching ratio from $t \bar t$ production at HL-LHC can be improved by an order of magnitude in $Zt$ production. Even stronger limits, at the $10^{-6}$ level for $tu$ and $10^{-5}$ level for $tc$ interactions, could be achieved at a future FCC-hh.

Besides the optimisation of the expected sensitivity, e.g. by varying the kinematical cuts, several improvements to the simple analyses presented here are possible, using additional information. In the semileptonic channel, the use of $b$-tagging would further reduce the main backgrounds from $Zj$ and $\gamma j$ production. In the hadronic channel, jet substructure observables like $N$-subjettiness~\cite{Thaler:2010tr} can be further used to reject the $Zj$ and $\gamma j$ backgrounds where the jets correspond to quarks and gluons. In both cases, the use of these variables requires a more delicate analysis, with the inclusion of pile-up and perhaps also a calibration of the Monte Carlo predictions against measured data. Such a study is beyond the scope of this work, and it is worth being pursued by experiments at the high luminosity LHC run. 

\section*{Acknowledgements}
This work has been supported by MINECO Projects  FPA 2016-78220-C3-1-P, FPA 2013-47836-C3-2-P (including ERDF), Junta de Andaluc\'{\i}a Project FQM-101 and  European Commission through the contract PITN-GA-2012-316704 (HIGGSTOOLS).


\begin{thebibliography}{99}

\bibitem{AguilarSaavedra:2004wm}
  J.~A.~Aguilar-Saavedra,
  Acta Phys.\ Polon.\ B {\bf 35} (2004) 2695
  [hep-ph/0409342].

\bibitem{Eilam:1990zc}
  G.~Eilam, J.~L.~Hewett and A.~Soni,
  Phys.\ Rev.\ D {\bf 44} (1991) 1473
   Erratum: [Phys.\ Rev.\ D {\bf 59} (1999) 039901].

\bibitem{Mele:1998ag}
  B.~Mele, S.~Petrarca and A.~Soddu,
  Phys.\ Lett.\ B {\bf 435} (1998) 401
  [hep-ph/9805498].

\bibitem{AguilarSaavedra:2002ns}
  J.~A.~Aguilar-Saavedra and B.~M.~Nobre,
  Phys.\ Lett.\ B {\bf 553} (2003) 251
  [hep-ph/0210360].

\bibitem{Han:1995pk}
  T.~Han, R.~D.~Peccei and X.~Zhang,
  Nucl.\ Phys.\ B {\bf 454} (1995) 527
  [hep-ph/9506461].

\bibitem{Han:1996ep}
  T.~Han, K.~Whisnant, B.~L.~Young and X.~Zhang,
  Phys.\ Rev.\ D {\bf 55} (1997) 7241
  [hep-ph/9603247].

\bibitem{Han:1996ce}
  T.~Han, K.~Whisnant, B.~L.~Young and X.~Zhang,
  Phys.\ Lett.\ B {\bf 385} (1996) 311
  [hep-ph/9606231].

\bibitem{AguilarSaavedra:2000aj}
  J.~A.~Aguilar-Saavedra and G.~C.~Branco,
  Phys.\ Lett.\ B {\bf 495} (2000) 347
  [hep-ph/0004190].

\bibitem{Hosch:1997gz}
  M.~Hosch, K.~Whisnant and B.~L.~Young,
  Phys.\ Rev.\ D {\bf 56} (1997) 5725
  [hep-ph/9703450].

\bibitem{Han:1998tp}
  T.~Han, M.~Hosch, K.~Whisnant, B.~L.~Young and X.~Zhang,
  Phys.\ Rev.\ D {\bf 58} (1998) 073008
  [hep-ph/9806486].

\bibitem{delAguila:1999ac}
  F.~del Aguila, J.~A.~Aguilar-Saavedra and L.~Ametller,
  Phys.\ Lett.\ B {\bf 462} (1999) 310
  [hep-ph/9906462].

\bibitem{Chatrchyan:2013nwa}
  S.~Chatrchyan {\it et al.} [CMS Collaboration],
  Phys.\ Rev.\ Lett.\  {\bf 112} (2014) no.17,  171802
  [arXiv:1312.4194 [hep-ex]].

\bibitem{Sirunyan:2017kkr}
  A.~M.~Sirunyan {\it et al.} [CMS Collaboration],
  JHEP {\bf 1707} (2017) 003
  [arXiv:1702.01404 [hep-ex]].

\bibitem{Aad:2015uza}
  G.~Aad {\it et al.} [ATLAS Collaboration],
  Eur.\ Phys.\ J.\ C {\bf 76} (2016) no.1,  12
  [arXiv:1508.05796 [hep-ex]].



\bibitem{ATL-PHYS-PUB-2016-019}
      ATLAS Collaboration,
note ATL-PHYS-PUB-2016-019.










\bibitem{delAguila:1999kfp} 
  F.~del Aguila and J.~A.~Aguilar-Saavedra,
  Nucl.\ Phys.\ B {\bf 576}, 56 (2000)
  [hep-ph/9909222].

\bibitem{AguilarSaavedra:2008zc}
  J.~A.~Aguilar-Saavedra,
  Nucl.\ Phys.\ B {\bf 812} (2009) 181
  [arXiv:0811.3842 [hep-ph]].

\bibitem{AguilarSaavedra:2010rx}
  J.~A.~Aguilar-Saavedra,
  Nucl.\ Phys.\ B {\bf 837} (2010) 122
  [arXiv:1003.3173 [hep-ph]].

\bibitem{protos}
J. A. Aguilar-Saavedra.
PROTOS, a PROgram for TOp Simulations. http://jaguilar.web.cern.ch/jaguilar/protos/




\bibitem{Sjostrand:2007gs}
  T.~Sjostrand, S.~Mrenna and P.~Z.~Skands,
  Comput.\ Phys.\ Commun.\  {\bf 178} (2008) 852
  [arXiv:0710.3820 [hep-ph]].

\bibitem{deFavereau:2013fsa}
  J.~de Favereau {\it et al.} [DELPHES 3 Collaboration],
  JHEP {\bf 1402} (2014) 057
  [arXiv:1307.6346 [hep-ex]].


\bibitem{Cacciari:2011ma}
  M.~Cacciari, G.~P.~Salam and G.~Soyez,
  Eur.\ Phys.\ J.\ C {\bf 72} (2012) 1896
  [arXiv:1111.6097 [hep-ph]].

\bibitem{Cacciari:2008gp}
  M.~Cacciari, G.~P.~Salam and G.~Soyez,
  JHEP {\bf 04} (2008) 063
  [arXiv:0802.1189 [hep-ph]].

\bibitem{Krohn:2009th}
  D.~Krohn, J.~Thaler and L.~T.~Wang,
  JHEP {\bf 1002} (2010) 084
  [arXiv:0912.1342 [hep-ph]].

\bibitem{Alwall:2014hca}
 J.~Alwall {\it et al.},
 JHEP {\bf 1407} (2014) 079
 [arXiv:1405.0301 [hep-ph]].

\bibitem{Mangano:2016jyj}
  M.~L.~Mangano {\it et al.},
  CERN Yellow Report (2017) no.3,  1
  [arXiv:1607.01831 [hep-ph]].

\bibitem{Brust:2014gia}
  C.~Brust, P.~Maksimovic, A.~Sady, P.~Saraswat, M.~T.~Walters and Y.~Xin,
  JHEP {\bf 1504} (2015) 079
  [arXiv:1410.0362 [hep-ph]].

\bibitem{Aguilar-Saavedra:2014iga}
  J.~A.~Aguilar-Saavedra, B.~Fuks and M.~L.~Mangano,
  Phys.\ Rev.\ D {\bf 91} (2015) 094021
  [arXiv:1412.6654 [hep-ph]].
  
\bibitem{Thaler:2010tr}
  J.~Thaler and K.~Van Tilburg,
  JHEP {\bf 1103} (2011) 015
  [arXiv:1011.2268 [hep-ph]].

\end{thebibliography}
\end{document}